\documentclass[prl,aps,twocolumn]{revtex4}
\usepackage{epsfig}
\usepackage{bm}

\usepackage[latin1]{inputenc}
\newcommand{\beq}{\begin{equation}}
\newcommand{\eeq}{\end{equation}}
\newcommand{\bea}{\begin{eqnarray}}
\newcommand{\eea}{\end{eqnarray}}
\begin{document}
\title{Non-universality  in Micro-branching Instabilities in Rapid Fracture: the Role of Material Properties}
\author{Eran Bouchbinder}
\author{Itamar Procaccia}
\affiliation{Dept. of Chemical Physics, the Weizmann Institute of Science, Rehovot 76100, Israel}
\begin{abstract}
In spite of the apparent similarity of micro-branching instabilities in different brittle materials, we propose that the physics determining the typical length- and time-scales characterizing the post-instability patterns differ greatly from material to material. We offer a scaling theory connecting the pattern
characteristics to material properties (like molecular weight) in brittle plastics like PMMA, and stress
the fundamental differences with patterns in glass which are crucially influenced by 3-dimensional dynamics. In both cases the present ab-initio theoretical models are still too far from
reality, disregarding some fundamental physics of the phenomena.
\end{abstract}
\maketitle

\begin{figure}[here]
\centering \epsfig{width=.45\textwidth,file=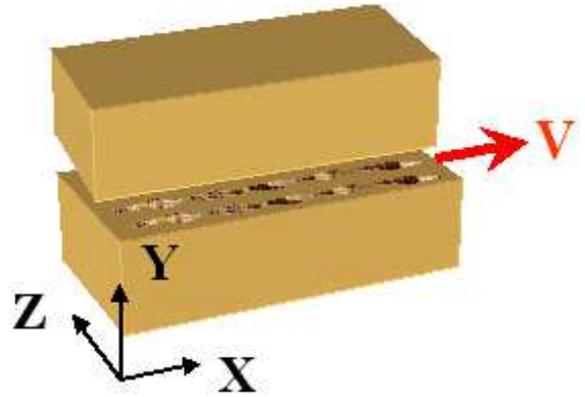}
\caption{Typical geometry of a dynamically generated crack, with the side branches that
result from the instability leaving their mark on the faces of the crack.} \label{geo}
\end{figure}

More than a decade of relatively high precision experiments on rapid
crack propagation in brittle materials has ignited the interest of
the physics community \cite{99MF} . It appears that one important
prediction of ``Linear Elasticity Fracture Mechanics" \cite{98Fre},
stating that cracks will accelerate smoothly as they lengthen until
they reach their asymptotic velocity (bounded either by the loading
conditions or by the Rayleigh speed $c_R$), is in fundamental
contradiction with experiments. In thin brittle plates of width $W$
(see the typical geometry in Fig. \ref{geo}; $W$ is in the $z$
direction, with length in $x$ and height in $y$ much larger than
$W$), at some finite fraction of the Rayleigh speed, the smooth
dynamics of the crack evolving in the $x$ direction is marred by the
appearance of micro-branches in the $x-y$ plane that do not cut
through the whole width $W$.  These micro-branches have a typical
length $\ell_b$, a typical width $\Delta z \leq W$ and they appear
to recur with an average periodicity. We will refer to this
recurrence scale as ``the scale of noisy periodicity" $\Delta x$.
The nature of this length-scale and the physics that determine it
are the subjects of this Letter. In glass this scale appears to
range between the measurement resolution ($\sim $1 $\mu$m) and a few
mm \cite{02SCF}, while in PMMA between tens and hundreds $\mu$m
\cite{96SF}, depending on the material properties and the
experimental conditions.
%%%%%%% FIGURE 1 %%%%%%%%%%%%%%%%%%
%%%%%%%%%%%%%%%%%%%%%%%%%%%%%%%%%%

As is well known, linear elasticity fracture mechanics has no typical scale, and thus the appearance of such a scale is yet another demonstration of the need to modify this theory in the context of dynamic fracture.  Indeed, there exist concepts in the classical theory of crack propagation that can be used to form a length scale.
Recall that in the classical theory one asserts that as a crack of length $L$ propagates at velocity $v$ under loading conditions $\sigma$, the energy $G(\sigma, L, v)$ released from the stressed material streams into the tip region where it is compensated by the total dissipation $\Gamma(v)$,
\begin{equation}
G(\sigma,L, v) =  \Gamma(v) \ . \label{Freund}
\end{equation}
The dissipation function $\Gamma(v)$ is not computable from elasticity theory, but it can be
measured in experiments, (essentially by using Eq. (\ref{Freund})). Equipped with this function
and the Young modulus $E$ of the material we can form a (velocity-dependent) scale  from
%%%%%%%%%%%%%
%%%%%%%%%%%%%%
\begin{equation}
\ell(v) \equiv \Gamma(v)/E \ . \label{scale}
\end{equation}
To get a rough estimate of the resulting scale, we use the data for PMMA determined in \cite{99MF} near the onset of the microbranching instability: $\Gamma(v_c) \simeq 3000 J/m^2$
and $E \simeq 3\cdot 10^{9} N/m^2$. This predicts $\ell(v)$ in the $\mu m$ range; a similar estimation for glass yields a $nm$ scale. We thus understand that the scale computed in Eq. (\ref{scale}) may
be relevant for PMMA, but appears utterly out of range for glass, where the observed scales of
noisy periodicity must be sought elsewhere, as we discuss below. Given information
about the largest stress value that marks the breakdown of the linear theory, say $\sigma_m$, another
length-scale can be constructed as
\begin{equation}
\tilde \ell(v) \equiv E \Gamma(v)/\sigma^2_m \ .
\label{tilscale}
\end{equation}
We estimate $\sigma_m$ by the yield stress of the material. For the materials under discussion the
resulting length-scale is typically two order of magnitude larger than (\ref{scale}), putting it right
in the range of scales of noisy periodicity for PMMA, but still off range for glass.
We continue now
to assess the relevance of (\ref{scale}) or (\ref{tilscale}) to PMMA.
Obviously, the dissipation $\Gamma(v)$ in PMMA should be a strong function of the molecular
weight $M$ of the polymers that make the material, due to the
increased density of entanglements of the polymer chains. Indeed, a
calculation of the $M$ dependence of $\Gamma$ in {\em quasi-static} conditions exists \cite{88MP}, with the resulting prediction
\begin{equation}
\Gamma(v=0) \sim e^{-M_0/M} \ , \label{Mdep}
\end{equation}
where $M_0$ is the smallest molecular weight below which the polymers do not contribute to $\Gamma$. On the other hand, the Young modulus $E$ and the yield stress were shown \cite{76KT,Vincent} to be $M$-independent.

The length (\ref{scale}) (or \ref{tilscale}) will be dressed by dynamical effects, which we introduce in the form
\begin{equation}
\ell(v) = e^{-M_0/M} f(v/c_R)\ . \label{vdep}
\end{equation}
To proceed, we adopt the scaling assumption that $\ell(v)$ is the
only typical scale in the problem, characterizing the length
$\ell_b$ of micro-branches in the $x-y$ plane as well as $\Delta x$
in the $x-z$ plane. In addition, we will assert that in PMMA (in
contrast to glass, and see below for further discussion) the
relevant length-scales are determined by a competition between the
side micro-branch and the main crack, without important dynamical
coupling to the third dimension. In this respect the mechanism
underlying the repetitive nature of the microbranching process in
PMMA is qualitatively similar to the 2-dimensional branching model
developed recently in \cite{05BMP}. In this model, the local
competition between the side branch and the main crack temporarily
slows the main crack below the relevant branching velocity. When the
main crack regains the critical velocity another side branching
event occurs (see Fig. 7 in \cite{05BMP}). It was found (see Fig. 6
in \cite{05BMP}) that the velocity dependence of $\ell_b$ has the
following form
\begin{equation}
\ell_b(v) = \ell_b(v_c)\left[1 + \alpha (v-v_c)/c_R\right] \  . \label{ellscale}
\end{equation}
Here $\ell_b(v_c)$ is the finite length of the side branch which occurs at threshold and
$\alpha$ is a dimensionless coefficient. In \cite{05BMP} this length
was introduced by hand to represent the scale over which linear elasticity theory is not
applicable (the so-called process zone); here we take it as the undressed length proportional to $\exp{(-M_0/M)}$. Collecting our assumptions together we present the prediction for the scale of noisy periodicity and the length of the micro-branches in PMMA:
\begin{equation}
\Delta x(v,M) \sim \ell_b(v,M) \sim
e^{-M_0/M}\left[1+\alpha\frac{v-v_c}{c_R}\right] \ .
\label{finalscale}
\end{equation}
Note that if indeed the scale $ \ell_b(v_c)$ is physically identified with the process zone,
it may have a velocity dependence of its own. If this velocity dependence is linear, the prediction
(\ref{finalscale}) remains unchanged.

\begin{figure}[here]
\centering \epsfig{width=.4\textwidth,file=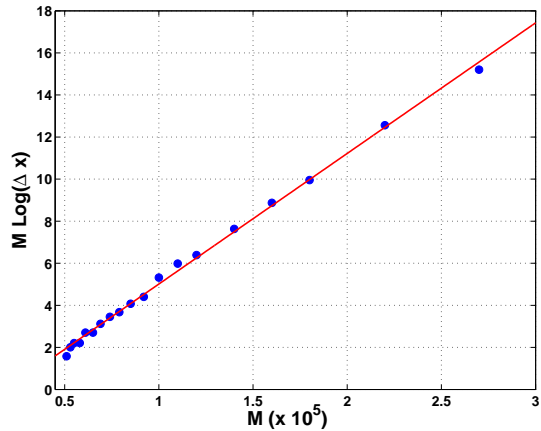}
\caption{$M~ log(\Delta x)$ as a function of $M$. The filled circles
are the experimental data taken from \cite{77KT} (see Table 1) and
the solid line is a linear fit.} \label{MW}
\end{figure}

\begin{figure}
\centering \epsfig{width=.4\textwidth,file=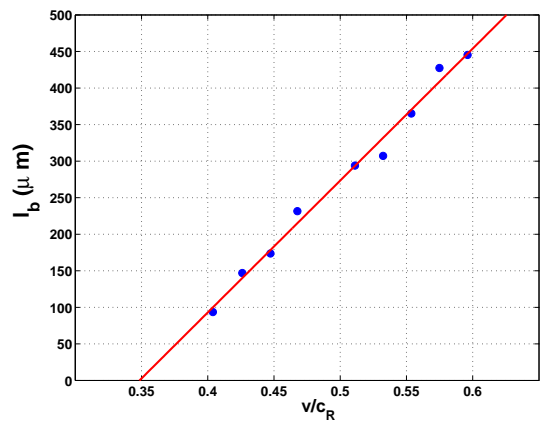}
\caption{The branch length $\ell_b$ as a function of the mean
normalized velocity for  PMMA. Data taken from \cite{98SF}. The
solid line represent the liner fit with $\alpha\approx 60$,
$v_c\approx 0.365 c_R$, and  $\ell(v_c)\approx 30 \mu$m.}
\label{BranchLength}
\end{figure}

It is possible to test this major prediction against available
experiments. The influence of the mean molecular weight $M$ on the scale of noisy
periodicity in PMMA was studied in \cite{77KT}  at
a given average velocity. Fig. \ref{MW} displays their experimental data
for $M~ log(\Delta x)$ as a function of $M$ (taken from Table 1 in
\cite{77KT}).
%%%%%%% FIGURE 2 %%%%%%%%%%%%%%%%%%
%%%%%%%%%%%%%%%%%%%%%%%%%%%%%%%%%%
The almost perfect linearity of the graph lends a strong support to the $M$
dependence in our scaling law. Next,
the branch length at a fixed mean molecular weight  as
a function of the normalized mean velocity for PMMA is shown in
Fig. \ref{BranchLength}. It is clear that $\ell_b$ is linear in
$v$ as predicted by our scaling relation. Moreover, the fit indicates
the existence of a finite length $\ell(v_c)\approx 30 \mu$m as predicted by the scaling law.
This finite length is determined uniquely in this graph once $v_c$ is known.
The value of $\ell(v_c)$ is confirmed in \cite{96SF}.
%%%%%%% FIGURE 3 %%%%%%%%%%%%%%%%%%
%%%%%%%%%%%%%%%%%%%%%%%%%%%%%%%%%%

In summary, the scale of noisy periodicity in PMMA appears to be
adequately described by a scaling theory in which the main scale
depends on the material properties $\Gamma$, $E$ and possibly
$\sigma_m$, dressed by dynamical effects which are not fundamentally
dependent on coupling to dynamics in the third dimension. There is
nothing like that in glass. First, we remarked before that the
length-scales (\ref{scale}) and (\ref{tilscale}) are in 1-100 nm
range, whereas the noisy periodicity in glass appears on scales up to a few mm.
Moreover, we will point out now that there is no typical
scale of noisy periodicity in glass experiments; the distances
between micro branching events can vary over three orders of
magnitude (from the measurement resolution to the scale of the width
$W$) and they are determined by a dynamical mechanism that rests
crucially on the dynamics of the crack front in the narrow third
dimension along the width $W$.

\begin{figure}[here]
\centering \epsfig{width=.4\textwidth,file=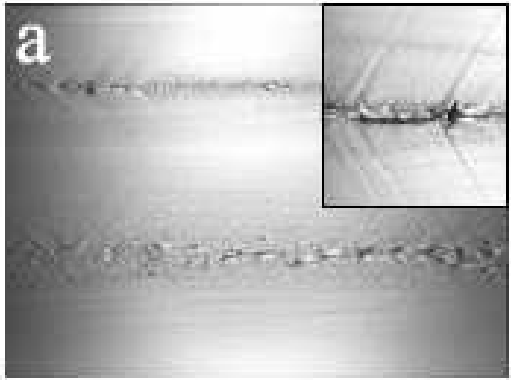}
\caption{The correlated micro-branches in glass samples. Each series
of micro-branches is localized in a strip of width $\Delta z$, and
has a noisy periodicity $\Delta x$$\simeq$4-5$\Delta z$. Adapted
from \cite{02SCF}.} \label{directed}
\end{figure}

\begin{figure}[here]
\epsfig{width=.45\textwidth,file=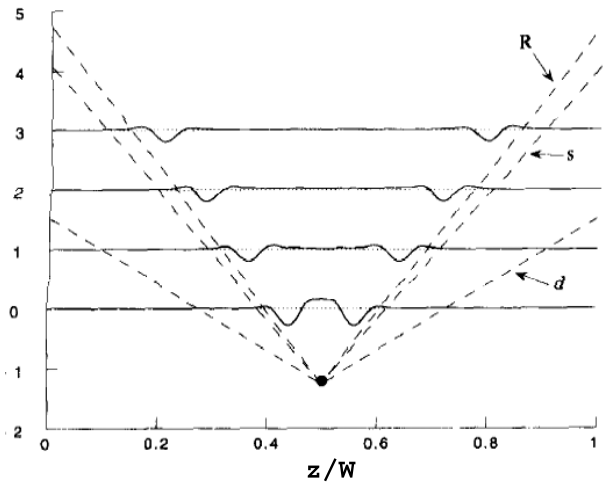}
\caption{The front waves
generated by the interaction of the front with an asperity. The
front velocity profile is plotted at constant time intervals. Note
the velocity overshooting just ahead of the asperity. $d,s,R$ are
the cones generated by the dilatational, shear and Rayleigh waves,
respectively. The relation $c_{FW} \approx c_R$ is apparent. Adapted
from \cite{98MR}.} \label{Rice}
\end{figure}

In glass,
once a micro-branch forms at a location $z$ along the
width of the sample and with a width $\Delta z$, the next branching
events  fall very precisely along a strip of width $\Delta z$
localized at the same location $z$; see Fig. \ref{directed} as an
example. The crucial phenomenon that is responsible for these highly
correlated structures is the interaction of the crack front with
spatial heterogeneities (regions of variable $\Gamma$). These can
generate non-decaying waves that travel with velocity $c_{FW}$
(relative to the heterogeneity) on the front as it propagates
\cite{97RF}. Plane numerical simulations \cite{98MR} have shown that
the interaction of the front with an asperity leads to a velocity
overshoot {\em exactly ahead} of the
interaction site, see Fig. \ref{Rice}.

%%%%%%% FIGURE 4 %%%%%%%%%%%%%%%%%%
%%%%%%%%%%%%%%%%%%%%%%%%%%%%%%%%%%

Since the creation of a micro-branch increases $\Gamma$ locally, one
expects the generation of front waves, as indeed was observed in
glass and can be seen in the inset of Fig. \ref{directed}. The
resulting velocity overshoot just ahead of the micro-branching
event was interpreted by Sharon, Cohen and Fineberg \cite{02SCF} as
the source of the well-defined lines of micro-branches in glass. In
this picture, the creation of a micro-branch reduces locally the
velocity of the main crack below the micro-branching velocity, but
this velocity is reached again due to the velocity overshoot just
ahead. The typical time $\tau$ for this process is the time needed for the
front wave to travel a distance of the order of the width of the
micro-branch $\Delta z$. Since the velocity of the front wave along
the front is $\sqrt{c^2_{FW}-v^2}$ one obtains
\begin{equation}
\tau \sim \frac{\Delta z(v)}{\sqrt{c^2_{FW}-v^2}} \ .
\label{tau_glass}
\end{equation}
Therefore, the noisy periodicity scale $\Delta x$ is estimated as
\begin{equation}
\Delta x(v) \sim \frac{\Delta z(v)~v}{\sqrt{c^2_{FW}-v^2}} \ .
\label{spacing_glass}
\end{equation}

Since the combination $v/\sqrt{c^2_{FW}-v^2}$ is a slowly increasing
function of $v$ ($0.94c_R$$\leq$$c_{FW}(v)$$\leq$$c_R$) in the
relevant range of velocities, one expects the ratio $\Delta
x(v)/\Delta z(v)$ to be nearly constant and this is indeed the case
as shown in \cite{02SCF}, where the relation is stated to be $\Delta
x\simeq$4-5$\Delta z$. The point that we want to stress here is that
in contradistinction with PMMA, in glass {\em there exist no typical
scale for the noisy periodicity}. $\Delta x(v)$ can be {\em
anything}, depending on the local asperity size that nucleated the
first micro branch of size $\Delta z(v)$. After that the dynamics
will sustain repeated micro-branches every $\Delta
x\simeq$4-5$\Delta z$. The only thing that can happen is that
$\Delta z(v)$ might increase as a function of $v$, which is the last
issue that we discuss.

%%%%%%% FIGURE 5 %%%%%%%%%%%%%%%%%%
%%%%%%%%%%%%%%%%%%%%%%%%%%%%%%%%%%

As noted earlier, Eq. \ref{Freund} predicts that cracks
accelerate smoothly towards their asymptotic velocity. A crucial
assumption leading to this prediction is that a single crack whose
front is essentially a point (there is translational invariance in
the z-axis) can expend the extra energy flowing to its tip by
adjusting its velocity and increasing the kinetic energy of its
surroundings. This assumption holds up to $v_c$, but breaks down for
$v>v_c$. At the onset of instability a {\em multiple} crack state is
formed by repetitive frustrated micro-branching events and these
micro-branches break the translational invariance in the third
dimension, occupying a finite width $\Delta z \leq W$. The essence
of the instability is that the additional energy supply is expended on creating
more surface area per unit crack extension, even at a constant
velocity. This process introduces new {\em dynamic variables} to the
problem: the number of coexisting branches $n$, the morphology of
the branches in the x-y planes and their width $\Delta z$.
All these variables contribute to the additional energy dissipation
per unit crack extension. It was shown \cite{96SGF} that the total
surface energy created per unit crack extension is proportional to
the total energy release rate, implying that in this regime the
fracture energy $\Gamma(v)$ is constant,  independent of the
velocity. Assuming that $\Delta z$ characterizes
the width of the micro-branches as long as they exist, we can write
\begin{equation}
W G(v) =  \Delta z(v) f(n(v), ...)\Gamma + W \Gamma \ ,
\label{NewBalance}
\end{equation}
where the right hand side is a sum of the dissipative contributions
from the micro-branches as well as from  the main crack, respectively. The
function $f(n(v), ...)$ depends on the number of coexisting
micro-branches, their morphology and maybe other
dynamic variables. The last relation can be rewritten as
\begin{equation}
\Delta z(v) =  f(n(v), ...) W\left( \frac{G(v)}{\Gamma} -1\right) \ .
\label{Del_z}
\end{equation}

\begin{figure}[here]
\centering \epsfig{width=.45\textwidth,file=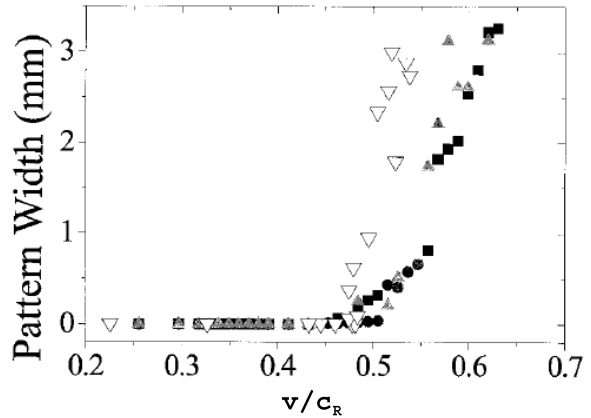}
\caption{The pattern width $\Delta z$ as a function of the
normalized mean velocity $v/c_R$ for PMMA (filled symbols) and glass
(empty symbols). Adapted from \cite{98SF}.} \label{delta_z}
\end{figure}

Although  $f(n(v), ...)$ is not known in adequate detail, making the last equation
somewhat tentative, we can still draw relevant qualitative conclusion.  Since $G(v)$
is known from experiments to be a strongly increasing function of $v$
\cite{96SGF}, we can safely expect that  $\Delta z(v)$ is also a strongly
increasing function of $v$. This expectation is supported by the data; the dependence of $\Delta z$ on the mean
velocity $v$ for both PMMA and glass is shown in Fig. \ref{delta_z}.
We stress that in glass the increase in
$\Delta z(v)$ {\em must} affect the scale of noisy periodicity, whereas in PMMA it does not.
In glass, the dynamics along the third dimension are
{\em strongly} coupled to the dynamics in the longitudinal direction
(by the front waves mechanism) and the dynamic length-scale $\Delta z$
controls all the other length-scales. In PMMA, the dynamics along
the third dimension are only {\em weakly} coupled to the dynamics in the
longitudinal direction, where the length-scales are determined by
material properties and the elastodynamic competition between the
micro-branches and the main crack.
%%%%%%% FIGURE 6 %%%%%%%%%%%%%%%%%%
%%%%%%%%%%%%%%%%%%%%%%%%%%%%%%%%%%

In summary, models of dynamic fracture that are developed based on the continuum equations supplemented
by criteria for crack evolution can hope to adequately describe the dynamic instabilities only if the physics
that determines the fundamental scales were incorporated. In glass it appears that without adequate treatment
of the third dimension and the physics of the crack front, not much insight into the micro-branching process can  be gained.
In PMMA the coupling to the third dimension might be weaker, but one should resolve how the material parameters appear in the continuum description. Completely different scaling laws appear in these two materials due to the different
mechanisms that control the scale of noisy periodicity; probably other materials
may increase the richness of the phenomenology. It might be worthwhile to stress less the presence of universal features
and more the interesting physics that can be learned even with simple scaling theories.

\end{document}